\documentclass{JHEP3}
\newcommand{\beq}{\begin{equation}}
\newcommand{\eeq}{\end{equation}}
\newcommand{\psib}{\ensuremath{\overline{\psi}}}
\usepackage{epsfig,multicol}

\title{Lattice Supersymmetry and Topological Field Theory}

\author{Simon Catterall\\
	Department of Physics, Syracuse University, Syracuse, NY 13244, USA\\
	E-mail: \email{smc@physics.syr.edu}}

\preprint{SU-4252-774}	

\abstract{
It is known that certain theories with extended supersymmetry can be
discretized in such a way as to preserve
an exact fermionic symmetry. In the simplest model of this kind,
we show that this residual supersymmetric invariance is
actually a BRST symmetry associated with gauge fixing an 
underlying local shift symmetry. Furthermore, the starting lattice action
is then seen to be entirely a gauge fixing term. The corresponding continuum
theory is known to be a topological field theory. We look, in detail,
at one example - supersymmetric quantum mechanics which possesses
two such BRST symmetries. 
In this case, we show that the lattice theory can be obtained by blocking
out of the continuum in a carefully chosen background
metric. 
Such a procedure will not change the Ward
identities corresponding to the BRST symmetries
since they correspond to topological observables. Thus, at the quantum
level, the continuum BRST symmetry is preserved in the lattice theory.
Similar conclusions are reached for the two-dimensional complex
Wess-Zumino model and imply that all the supersymmetric Ward
identities are satisfied {\it exactly} on the lattice. Numerical results
supporting these conclusions are presented.
}


\begin{document} 

\section{Introduction}

Supersymmetric field theories are interesting both from phenomenological
and theoretical points of view - their improved U.V behavior offers the
hope of resolving the gauge hierarchy problem and they arise naturally
as low energy limits of string and M-theory.
Most of the interesting physics of such theories lies in non-perturbative
regimes. Discretization on a space-time lattice appears to provide a
natural way to study such theories and considerable effort has gone
into formulating such lattice supersymmetric theories \cite{gen}.
Unfortunately supersymmetry is typically broken at the classical
level in such models. At the quantum level the absence of such a symmetry
leads to an effective action containing relevant SUSY-violating 
interactions. To achieve a supersymmetric continuum limit then 
necessitates fine tuning the couplings to each of these
operators. In most situations this is prohibitively difficult.

In an previous paper \cite{wz2}
we advocated a different approach -- try to preserve
a subset of the full supersymmetry in the lattice model. 
A similar approach has also been adopted in \cite{kaplan}.
Numerical simulations of two models where this idea can
be implemented explicitly lend strong support to the idea
that preserving some subset of the continuum supersymmetry
transformations
can indeed protect the lattice model from the 
dangerous radiative corrections that generically
plague discretizations of supersymmetric field theories \cite{wz2,qm}.
In this paper we offer a proof of this statement
for the case of supersymmetric quantum mechanics and the
two-dimensional complex Wess-Zumino model. Indeed, our
result is much stronger - by careful choice of the lattice
action we can
show that there are {\it no} cut-off effects in the 
lattice supersymmetric Ward identities even for 
supersymmetries which are
broken at the classical level.

We will argue that the reason 
that these lattice models are well behaved 
is that they
are related to Witten-type continuum
topological field theories \cite{rev}.
Such theories are constructed using a nilpotent
symmetry formed from elements of the
original supersymmetry. A key feature of such theories is that
they contain observables whose expectation values are independent
of the metric. After the partition function itself, the simplest
examples of such observables are the
Ward identities corresponding to the nilpotent
symmetry. This metric independence plays a crucial
role in allowing us to establish a direct link between
the continuum and lattice theories and allows the latter
to possess rather remarkable properties.

In the next section we introduce a simple lattice model 
which exhibits an exact fermionic symmetry and show that
this symmetry is actually a BRST symmetry following from
fixing a local topological symmetry. We then show how
this topological model can be used to describe
a lattice regularized theory of supersymmetric
quantum mechanics \cite{qm}. In this latter case we show that
the model is naturally associated with two
independent BRST symmetries. In the continuum these two symmetries
exhaust the original supersymmetries and show that  
that supersymmetric quantum mechanics can indeed be viewed as
a topological field theory. 

For convenience we include a brief summary of
the main features of
such topological field theories. We then show how
this continuum topological structure 
allows us to derive the lattice theory 
by integrating out the continuum fields in a carefully
chosen background geometry. More precisely, we consider
metrics which are continuous functions of a parameter $\beta$ such
that in the limit $\beta\to\infty$ a ``lattice'' structure is
induced in the model. For any finite value of $\beta$ we can perform
an associated  $\beta$-dependent change of variables in the continuum model
which preserves the topological symmetries. Hence such a
procedure ensures that any topological observable is {\it independent}
of $\beta$. We show that indeed, a local, lattice theory is approached in the limit
of large $\beta$. Furthermore, our construction then guarantees 
that the resulting lattice theory retains an element of
supersymmetry at the quantum level.

The same
analysis applied to the complex Wess Zumino model in two dimensions allows
us to write down lattice actions with respect to which
all supersymmetric Ward identities are satisfied exactly. Numerical
results confirming these conclusions are presented.
Finally
we discuss the prospects for extending these ideas to
more realistic models.

\section{Simple Example}

Consider the simple model discussed in \cite{wz2}
\begin{equation}
S=\frac{1}{2\alpha}N_i^2(x)+\overline{\psi}_i\frac{\partial N_i}{\partial x_j}\psi_j
\label{eq1}
\end{equation}
which admits a fermionic symmetry
\begin{eqnarray*}
\delta x_i &=& \psi_i\xi \nonumber\\
\delta \overline{\psi}_i &=& N_i\xi \nonumber \\
\delta \psi_i &=& 0 \nonumber
\end{eqnarray*}
Here, $N_i(x)$ is an {\it arbitrary} function of the scalar field $x_i$ and
$\psi_i$, $\overline{\psi}_i$ are real, independent grassmann variables. Notice that
this transformation is nilpotent on-shell. Indeed, if we introduce
an auxiliary (commuting) field $B_i$ we can define a new action
\begin{equation}
S^\prime=-\frac{1}{2}\alpha B_i^2+N_iB_i+\overline{\psi}_i\frac{\partial N_i}{\partial x_j}\psi_j
\label{act}
\end{equation}
If we integrate over $B_i$ (along the imaginary axis)
we just recover the original action $S$. This
new action also has an invariance
\begin{eqnarray*}
\delta_{(1)} x_i &=& \psi_i\xi \nonumber\\
\delta_{(1)} \overline{\psi}_i &=& B_i\xi \nonumber \\
\delta_{(1)} \psi_i &=& 0 \nonumber\\
\delta_{(1)} B_i &=& 0
\end{eqnarray*}
The significance of the subscript labeling the action and symmetry
variation
will be become apparent later.
It is easy to see that this new transformation is nilpotent off-shell
$\delta_{(1)}^2\Phi=0$ for any of the fields $\Phi=\{\psi,\psib,x,B\}$. 
More importantly it is clear 
that the new action $S^\prime$ is nothing but the variation of
another function
\begin{displaymath}
S^\prime\xi=\delta_{(1)}\left(\overline{\psi}_i\left(N_i-\frac{1}{2}\alpha B_i\right)\right)
\end{displaymath}
Thus we recognize our original invariance as a BRST invariance and our
original action as nothing but a gauge fixing term! 
The topological origins of
the lattice theory are made more
manifest when it is realized that the local gauge symmetry
which is being fixed to generate the
BRST invariance is nothing but an arbitrary shift in the
scalar field $x_i$. 
Imagine a classical theory depending on a scalar field $x_i$ with trivial
classical action $S(x)=0$. Clearly this theory is invariant under a huge
local symmetry - namely arbitrary shifts in the scalar
field 
\begin{displaymath}
x_i\to x_i+\epsilon_i
\end{displaymath}
To quantize we need to pick a gauge. One simple way
to do is is to impose $N_i(x)=0$ where $N_i(x)$ is some arbitrary function
of the field $x_i$. Then the partition function will be
\begin{displaymath}
Z=\int dx_i\, \delta(N_i) {\rm det}\left(\frac{\partial N_i}{\partial x_j}\right)
\end{displaymath}
If we represent the determinant using anticommuting ghosts and introduce
a multiplier field $B_i$ for the delta function we
recover our simple model eqn.~\ref{eq1} in $\alpha=0$ gauge. The
usual theorem associated with quantization of gauge theories allows us
to relax this Landau-like gauge to a Feynman-like gauge with
$\alpha$ non-zero without changing the expectation values of
gauge invariant quantities.
Notice that the physical `fermions' of the SUSY theory are to be
identified with the ghosts of the gauge fixed scalar field theory.

There is another simple way to see the the model written down in eqn.~\ref{eq1}
exhibits an unusual symmetry. If we imagine performing a
change of variables in the path integral defining this theory according
to $\eta_i=N_i(x)$ the Jacobian of this transformation cancels
the fermionic determinant and the partition function factorizes
into a product of gaussians \cite{nicolai}
\begin{displaymath}
Z=\prod_i d\eta_i e^{-\frac{\eta^2}{2\alpha}}
\end{displaymath}
This resultant partition function is trivially invariant under the same local
shift symmetry discussed earlier. A transformation of this type which
cancels off the fermion determinant is called a Nicolai map and it should now be
clear that the existence of a {\it local} Nicolai map can
be attributed to the presence of a topological symmetry \cite{rev}.
 
We now turn to the simplest application of these ideas - supersymmetric
quantum mechanics

\section{Supersymmetric Quantum Mechanics}
\label{susyqm}

Imagine now choosing the
function
$N_i(x)=N^{(1)}_i(x)$ corresponding to an action $S^\prime=S_{(1)}$ where 
\begin{displaymath}
N^{(1)}_i(x)=D^S_{ij}x_j+P^\prime_i(x)
\end{displaymath}
In this expression $D^S$ represents the symmetric difference operator
and
$P^\prime_i(x)$ is some arbitrary polynomial in the lattice
field $x_i$. To be concrete we can imagine a model with a single
interaction coupling $g$
of the form
\begin{equation}
P^\prime=mx_i+m^W_{ij}x_j+gx_i^3
\label{pot}
\end{equation}
Notice that we are also free to add a Wilson  mass term $m^W$ to the potential
to eliminate lattice doubles associated with the choice of lattice
derivative operator $D^S$.

In this case we recognize our simple model as a lattice regularized version
of supersymmetric quantum mechanics - a model well known to
possess  
a topological field theory interpretation \cite{rev}. We discuss some of
the generic features of such theories in the next section
but suffice it to say here that such theories possess
observables whose expectation values are metric
independent. Clearly, on the
lattice, there is no notion of continuum metric and
so in a strict sense the lattice model cannot be said to
be topological. However, the fact that the action is a BRST
variation of a local function of the lattice fields clearly
imposes strong restrictions on the quantum theory. Indeed we shall
see that in this model certain symmetries which are broken at the
level of the classical lattice action are restored in the full quantum theory.

Notice also that the partition function of this model (in $\alpha=0$ gauge)
just reduces to
an integral over the set of field
configurations satisfying the gauge condition. In the case of
supersymmetric quantum mechanics this is just
the moduli space of classical solutions to the
equation of motion.
On a circle these classical solutions are just a finite set
of points
$x_i=x_c$ where $P^\prime(x_c)=0$. Thus the partition
function just reduces to a sum over the critical points of the
potential $P(x)$.
Notice that this
solution is {\it independent} of the lattice cut-off -- indeed it is
the same result one would have gotten for the analogous continuum
model. 

Furthermore we can consider another gauge condition which corresponds
to the same classical moduli space
$N_i(x)=N^{(2)}_i(x)$ with
\begin{displaymath}
N^{(2)}_i(x)=D^S_{ij}x_j-P^\prime(x)
\end{displaymath}
We can construct the action $S_{(2)}$ following from this
gauge condition by the same procedure and furthermore
after exchanging the roles of ghost $\psi$ and antighost
$\overline{\psi}$ we can easily see it only differs from $S_{(1)}$ by
the addition of a simple operator $C(g)$ 
\begin{displaymath}
S_{(2)}=S_{(1)}-2C
\end{displaymath}
The operator $C=D^S_{ij}x_jP^\prime_i(x)$ and corresponds to the integral
of a 
total derivative term in the continuum. On the lattice it will be non-zero
if $P^\prime(x)$ contains nonlinear powers of the field $x$ which is the
case for an interacting model with $g\ne 0$.

The variation of the fields under this second (nilpotent) BRST
symmetry is
\begin{eqnarray*}
\delta_{(2)} x_i &=& \xi\overline{\psi}_i \nonumber\\
\delta_{(2)} \psi_i &=& \xi\left(B_i-2P^\prime_i(x)\right) \nonumber \\
\delta_{(2)} \overline{\psi}_i &=& 0 \nonumber\\
\delta_{(2)} B_i &=& 2\xi P^{\prime\prime}_i\psib
\end{eqnarray*} 
In the continuum where $S_{(1)}=S_{(2)}$ the action of supersymmetric
quantum mechanics would 
then possess {\it two} BRST
invariances. On the lattice if we choose $S_{(1)}$ as action we no longer
have $\delta_{(2)}$ as a symmetry (except for a free theory)
and vice versa. Thus at the {\it classical} level
discretization on a lattice necessarily breaks one of the continuum
symmetries. However, we will see 
that this symmetry is {\it restored} at the quantum level with 
lattice Ward identities corresponding to both $\delta_{(1)}$ and
$\delta_{(2)}$ being satisfied exactly for
arbitrary lattice spacing. Indeed we will prove that
there exists a one parameter family of lattice actions in which all
of the BRST Ward identities are satisfied with {\it no} cut-off
effects. This feature is crucially dependent on the existence of
this topological symmetry. 
 
\section{Continuum Topological Field Theories}

In this section we give a brief review of some of the general features
of continuum topological field theories \cite{rev,rev2}. Such theories are formulated
on a n-dimensional manifold equipped with a metric $g_{\mu\nu}$. On this
space there exists a set of fields $\Phi$ and an
action $S\left(\Phi\right)$. Typically these theories contain
operators or topological observables $O_{\beta}\left(\Phi\right)$ with
the property that their expectation values are metric independent
\begin{displaymath}
\frac{\delta}{\delta g^{\mu\nu}}\left<O_{\beta_1}\ldots O_{\beta_2}\right>=0
\end{displaymath}
One way to guarantee this corresponds to the case in which there exists
a nilpotent symmetry $\delta$ such that
\begin{displaymath}
\delta O_\beta =0, \;\;\;\;\; T_{\mu\nu}=\delta G_{\mu\nu}
\end{displaymath}
The energy-momentum tensor $T_{\mu\nu}=\frac{\delta}{\delta g^{\mu\nu}}
S(\Phi)$. 
These conditions lead to the following expressions for
the expectation values of topological variables.
\begin{displaymath}
\frac{\delta}{\delta g^{\mu\nu}}\left<O_{\beta_1}\ldots O_{\beta_2}\right>=
-\int D\Phi \delta\left(
O_{\beta_1}\ldots O_{\beta_2}G_{\mu\nu}e^{-S(\Phi)}\right)=0
\end{displaymath}
Here we have also assumed that the measure is
invariant under the nilpotent symmetry
and that the observables themselves do not contain the
metric explicitly. Theories constructed in this way are called Witten type
or cohomological topological quantum field theories.  Typically the
nilpotent symmetry $\delta$ is realized as a BRST symmetry arising
from gauge fixing some underlying local shift symmetry.
In such theories
\begin{displaymath}
S\left(\Phi\right)=\delta \Lambda(\Phi)
\end{displaymath}
This latter result is very important as it guarantees that topological
observables and the partition function itself can be computed {\it exactly}
in the semi-classical limit. To see this introduce a parameter $\epsilon$
playing the role
of Planck's constant in the definition of a topological expectation
value and examine the variation of that expectation value
under variation of $\epsilon$
\begin{displaymath}
\frac{\partial}{\partial\epsilon}
\left<O_{\beta_1}\ldots O_{\beta_2}\right>=
\frac{1}{\epsilon^2}\int D\Phi \delta\left(O_{\beta_1}\ldots
O_{\beta_2}\Lambda(\Phi)\right)e^{-\frac{1}{\epsilon}S}=0
\end{displaymath}
Thus topological observables may be computed exactly
in the semi-classical approximation $\epsilon\to 0$. 
This semi-classical exactness may be translated into an independence
of topological observables on coupling. 
To see this consider an action of the form
\begin{displaymath}
S=S_0\left(\Phi\right)+g\Phi^n
\end{displaymath}
where the quadratic terms are contained in $S_0\left(\Phi\right)$ and we
allow for a generic interaction term. If we rescale the fields using
$\Phi\to\sqrt{\epsilon}\Phi$ it is easy to see that topological
observables computed in an ensemble with Planck constant
$\epsilon$ and coupling $g$ is equivalent to the same observable
computed in the ensemble $\epsilon=1$ and $g^\prime=g\epsilon^{n/2-1}$ 
The limit $\epsilon\to 0$ now corresponds to $g^\prime\to 0$ in the
latter ensemble and
hence the expectation value can be computed in the free field limit. 
For the case $O=1$ this implies that the partition function itself is
independent of $g$.
This, of course, is also
the property enjoyed by models with a local Nicolai map and makes
plausible the conjecture that models possessing a local Nicolai map
contain within them a topological symmetry.
  
A trivial set of topological observables correspond to operators of
the form
\begin{displaymath}
O=\delta O^\prime
\end{displaymath}
While trivial in a true topological sense (their expectation value
vanishes trivially on account of the nilpotent nature of $\delta$) they
will be of crucial importance in the parent
supersymmetric theory since they yield the supersymmetric
Ward identities.

\section{Relation Between Lattice and Continuum Theories}
\label{formal}

In section~\ref{susyqm} we discussed a lattice theory of supersymmetric
quantum mechanics and showed that it was possible to choose an
action which reproduced the continuum partition function exactly
(up to a parameter independent
multiplicative constant).
The simplest way to see
this utilizes the gauge $\alpha=0$ in which the partition
function reduces to a sum over the critical points of the
potential $P^\prime(x)=0$. The solution of this equation is 
identical in both lattice and continuum theories. Actually, the gauge
$\alpha=0$ can be used for an arbitrary topological observable
and implies that the expectation values for all such observables
only depend on the properties of the classical solution to the
field equations. This prompts us to guess that it should be
possible to forge an explicit
connection between the lattice and continuum theories 
useful for the computation of such observables. We will now
show that indeed this is the case.

One standard way to relate a lattice theory to
an underlying continuum theory derives from the
renormalization group. The lattice field at some point
is constructed by
averaging the continuum field over a neighborhood of that
point. This averaging or {\it blocking} procedure may be accomplished by
convolving the continuum field with a blocking function. Typically,
the precise shape of the blocking function is not important for
long distance physics.
The lattice or {\it block}
field which results from this operation is usually a
non-local function of the continuum fields. However, this
need not be the case for a topological field theory.
If we are
only concerned with the computation of topological observables we
are at liberty to block the continuum fields in an {\it arbitrary}
background metric. If this metric is then chosen carefully we can arrange for
the block fields to be related to the
continuum fields in a {\it completely local} manner.

Let us examine this in the case
of supersymmetric quantum mechanics. The bosonic part of
the continuum action for an arbitrary one-dimensional
metric written in terms of the einbein $e(t)$ takes the form (we have
integrated out the auxiliary field $B(t)$)
\begin{displaymath}
S=\int dt\, e(t) \left[\frac{1}{e(t)}\frac{dx}{dt}+P^\prime(x)\right]^2
\end{displaymath} 
Define now a scalar block
field $x^B(t)$ {\it in the continuum} as a convolution over the
original field $x(t)$ using a blocking function $B_\beta^-(t)$
\begin{equation}
x^B(t)=\int dt^\prime\, e(t^\prime)B_\beta^-\left(t-t^\prime\right)x(t^\prime)
\label{block}
\end{equation}
We will choose the blocking function 
$B_\beta^-(t)$ to be
\begin{displaymath}
B_\beta^-(t)=\frac{1}{2a}\left[L_\beta\left(t+\delta\right)-
L_\beta\left(t-a+\delta\right)\right]
\end{displaymath}
We require that the function $L_\beta(t)$ tend to the step function for
large $\beta$
\begin{displaymath}
\lim_{\beta\to\infty}L_\beta(t)= \theta(t)
\end{displaymath}
This choice of $B_\beta^-(t)$ ensures 
that for large $\beta$ the blocked field at point $t$
contains contributions from all points within a cell defined by
$-\delta \le t \le (a-\delta)$. Furthermore, we will require the
parameter $\delta\to 0^+$ at the end of the calculation.
A concrete example of such a function is given by $L_\beta(t)=\tanh{\beta t}$.
To capture the structure of the lattice
theory we will choose an associated ``lattice'' metric given by
$e(t)=e_\beta(t)$ where 
\begin{displaymath}
e_\beta(t)=\sum_{n=1}^N \frac{a}{A(\beta)} L_\beta^\prime(t-na)
\end{displaymath}
where the sum
runs over a finite set of $N$ points with ``lattice spacing'' $a$. 
The constant $A(\beta)$ is chosen so that 
\begin{displaymath}
\int dt e_\beta(t)=Na
\end{displaymath}
and we assume the continuum
theory is defined over a circle with circumference $Na$.
Notice that
\begin{displaymath}
\lim_{\beta\to\infty}L_\beta^\prime=\lim_{\beta\to\infty}\frac{dL_\beta}{dt}=
A_L\delta(t)
\end{displaymath}
where $A_L=A(\infty)$ is just some numerical coefficient.  
Returning to eqn.~\ref{block} we can now compute the block field
explicitly
\begin{displaymath}
x^B(t)\simeq \sum_{n=1}^N \frac{a}{A(\beta)} x(na) B_\beta^-(t-na)\;\; {\rm for\;large}\;\beta
\end{displaymath}
In the limit $\beta\to\infty$ this relation yields the result
\begin{displaymath}
\lim_{\beta\to\infty} x^B(t)= \sum_{n=1}^N
x(na)\frac{1}{2A_L}\left[\theta(t-na+\delta)-\theta(t-(n+1)a+\delta)\right]
\end{displaymath}
That is, the continuum
block field $x^B(t)$ is constant within each unit cell of the lattice
changing its value only on passing from one cell to the next. 
To proceed further it is necessary to compute its derivative.
\begin{eqnarray*}
\frac{dx^B(t)}{dt}&\simeq &\sum_{n=1}^N \frac{a}{A(\beta)} x(na) \frac{dB_\beta^-}{dt}(t-na)\nonumber\\
               &\simeq &\frac{1}{2A(\beta)}\sum_{n=1}^N
	       x(na)\left[L_\beta^\prime(t-na+\delta)-
	         L_\beta^\prime(t-(n+1)a+\delta)\right]
\end{eqnarray*}
Notice that this derivative does indeed vanish in the limit $\beta\to\infty$
for any point within a cell. To compute the action evaluated on
a block configuration in this
background we need to compute $\frac{1}{e_\beta}\frac{dx^B}{dt}$.
For any point within the cell $-\delta +na \le t \le (n+1)a-\delta$ the
leading contribution at large $\beta$ is seen to be
\begin{displaymath}
\frac{1}{e_\beta}\frac{dx^B}{dt}\simeq \frac{1}{2aA(\beta)}\left( f_\beta(z)
     \left[2x(na)-x((n+1)a)-x((n-1)a)\right]+
     \left[x((n+1)a)-x((n-1)a)\right]\right)
\end{displaymath}
where
\begin{displaymath}
f_\beta(z)=\left(\frac{L^\prime_\beta(a-z)-L^\prime_\beta(z)}    
     {L^\prime_\beta(z)+L^\prime_\beta(a-z)}\right)
\end{displaymath}
with $z=t-na$ and we have set $\delta$ to zero for simplicity.
This in turn reduces to
\begin{displaymath}
\lim_{\beta\to\infty}\frac{1}{e_\beta(t)}\frac{dx^B}{dt}=\sum_{n=1}^N
\frac{1}{2aA_L}\left[x(na)-x((n-1)a)\right]\left[\theta(t-(n-1/2)a)-\theta(t-(n+1/2)a)\right]
\end{displaymath}
Thus the derivative of a block function in such a background geometry
at large $\beta$ is constant now within a unit cell of the {\it dual
lattice}. These properties allow us to compute the integral
of an arbitrary function
of the block field $x^B(t)$ and its first derivatives. 
An example is the bosonic part of the continuum action $S_B$. This becomes  
\beq
\lim_{\beta\to\infty}S_B=\sum_{n=1}^N a
\left[\frac{1}{2A_La}D^-_{nm}x^B_m+P^\prime_n(x^B)\right]^2
\label{block_action}
\eeq
where the bosonic action now {\it only depends on the blocked fields
at the lattice points} and we use the obvious notation $x^B(na)\equiv x^B_n$.
The most striking thing about this block action $S_B$ is that it coincides with
the bosonic part of the lattice action  $S_{(1)}$ discussed earlier
if we identify the lattice field as the blocked continuum field
evaluated on a lattice point\footnote{a finite rescaling of the field is also
needed to make the correspondence exact - but this in turn is equivalent
to a rescaling of the coupling which will not affect topological observables}.
Notice also that the lattice theory we arrive at in this manner 
automatically incorporates an $r=1$ Wilson mass term to remove
potential lattice doublers (since $D^S-m_W=D^-$). 

Notice that the backward difference operator $D^-$ arises directly
from our choice of blocking function $B_\beta^-$. If we had made the
equally valid choice
\begin{displaymath}
B_\beta^+=\frac{1}{2a}\left[L_\beta\left(x+a-\delta\right)-
L_\beta\left(x-\delta\right)\right]
\end{displaymath}
we would have arrived at a block action of the same form as in
eqn.~\ref{block_action} but with $D^-$ replaced by the forward derivative
$D^+$.  

Generalization to
include the fermionic sector is straightforward and leads to
the conclusion that the total action when evaluated on block
configurations as defined by eqn.~\ref{block} goes over into
the full lattice action $S_{(1)}$ described in section~\ref{susyqm}.

This similarity between the blocked continuum theory at
large $\beta$ and the
lattice model described in section~\ref{susyqm} is intriguing but
by itself does not yet guarantee the exact equivalence of
continuum and lattice theories.
To complete
the correspondence between continuum and lattice
theories, and to show that for topological
observables the continuum expectation values are
just equal to their lattice counterparts, we still need to
argue that {\it only} block fields need to be taken into
account inside continuum path integrals as $\beta\to\infty$.
We offer two arguments for this.  

First consider the continuum boson kinetic term in terms of
the original fields
\begin{displaymath}
S_{\rm K}=\int dt \frac{1}{e_\beta}\left(\frac{dx}{dt}\right)^2
\end{displaymath}
As $\beta\to\infty$ it is clear that away from the lattice points
$t=na$ the action starts to diverge due to the presence of $e_\beta(t)$
in the denominator. Furthermore, 
it is clear that this effect yields an exponential
suppression of any field configuration $x(t)$ in which the field
changes rapidly within a cell. Indeed, we can argue that
that the only configurations that survive in the path integral
are those in which $\frac{dx}{dt}\sim e^{-\beta \delta t/2}$ for $\delta t$ away from
a lattice point. Thus, in the limit $\beta\to\infty$ only the block boson
fields survive in the path integral. The relations $\delta_{(1)} x=\psi$
and $\delta_{(2)} x=\psib$ then ensure that, in the absence
of topological symmetry breaking, only block fermion fields need
to be considered for large $\beta$.

Our second argument is formal but more general. We are
at liberty to  
consider the blocking transformation given in eqn.~\ref{block}
for {\it any} finite $\beta$ as corresponding 
to a simple change of variables in the continuum partition
function. 
Let us write this transformation for a generic field
$\Phi=\{ x,B,\psi,\psib \}$ using a compact notation
\begin{displaymath}
\Phi^B=K_\beta\Phi
\end{displaymath}
where the kernel $K_\beta$ is shorthand for
\begin{displaymath}
K_\beta\left(t,t^\prime\right)=e_\beta(t)B_\beta^-\left(t-t^\prime\right)
\end{displaymath}
and we leave implicit the integrals defining the convolution of $K_\beta$
with any field $\Phi$. 
In terms of these new variables the partition function
becomes
\begin{displaymath}
Z=\int D\Phi^B J(\beta) e^{-S\left(K^{-1}_\beta \Phi^B\right)}
\end{displaymath}
where the Jacobian $J(\beta)$ is {\it independent} of the fields since
the transformation is linear. 
Indeed, the linearity of this mapping implies that each of the 
BRST transformations for the original continuum
fields yields identical transformations of
the block fields. For example,
\begin{eqnarray*}
\delta_{(1)} x^B &=& \psi^B\xi \nonumber\\
\delta_{(1)} \overline{\psi}^B &=& B^B\xi \nonumber \\
\delta_{(1)} \psi^B &=& 0 \nonumber\\
\delta_{(1)} B^B &=& 0
\end{eqnarray*}
Thus the BRST operators remain nilpotent on the block fields. 
We assume that the inverse operator $K_\beta^{-1}$
exists.
For generic values of $\beta$ the resulting
action $S\left(K^{-1}_\beta \Phi^B\right)$ will be a complicated and non-local
function of the block fields. However, it has one crucial property -- it
will still be invariant under a BRST variation of the block fields. Indeed,
we can construct topological observables from the block fields just
as before 
\beq 
O_{\rm top.\; obs.}^B=\delta_{(i)}T^B(\Phi^B)
\label{obs}
\eeq 
for any function $T(\Phi^B)$ and either of the symmetries $i=1,2$.
Now imagine taking the limit $\beta\to\infty$. By examining the
form of the kernel it is easy to see that in this limit the block field becomes
an eigenvector of the kernel
\begin{displaymath}
\lim_{\beta\to\infty}K\Phi^B=\frac{1}{2A_L}\Phi^B
\end{displaymath}
Actually we must be careful here -- as $\beta\to\infty$ it is clear
that the operator $K$ develops a set of zero modes $f_l$ where
$Kf_l=0$ corresponding
to functions which are zero at the lattice points but are otherwise
unrestricted. These can be safely ignored only by considering the
form of the boson kinetic term which ensures that they are wholly
suppressed in the limit $\beta\to\infty$ as we argued above. Ignoring
these modes (which are not present at finite $\beta$) 
leads us to conclude that $K^{-1}_\beta \Phi^B\to 2A_L \Phi_B$ and thus
the action need only be evaluated for the block fields
\begin{displaymath}
\lim_{\beta\to\infty}S\left(K^{-1}_\beta \Phi^B\right)=S\left(2A_L \Phi^B\right)
\end{displaymath}

Thus in this limit the partition function factorizes into a piece
determined by the block fields at a finite set of ``lattice''
points labeled
by an integer $n$ together with an integral over all points lying in
cells of the ``lattice''
\begin{displaymath}
\lim_{\beta\to\infty}Z=J(\infty)\int D\Phi^B_{\rm cells} \int D\Phi^B_n e^{-S(2A_L \Phi^B_n)} 
\end{displaymath}
Throwing away this irrelevant (infinite) multiplicative constant
we see that we have arrived at our original lattice action
determined by a finite set of variables!
It is not hard to show that in the large $\beta$ limit
the observables given in eqn.~\ref{obs} go over into the
lattice Ward identities. Since our lattice theory has been
obtained by a process of continuous deformation of the continuum
theory we expect the resultant lattice theory to preserve the
values of {\it all} continuum topological observables.
Notice though that at $\beta=\infty$ the continuum block fields are
discontinuous functions. Thus we should not be surprised if
the Leibniz rule fails when applied to functions of
such fields and seeming total derivative
terms do not vanish in the block action. 

In the language of the renormalization group we have found that
the lattice action of section~\ref{susyqm} is a {\it perfect}
lattice action for the computation of topological observables -- it
yields cut-off independent predictions for the
corresponding expectation values. Thus the lattice theory
should contain a set of exact Ward identities corresponding
to any combination of supersymmetries which yields 
a continuum topological symmetry. The arguments presented above,
while not rigorous, are well supported by the numerical
results, as we will show in the next section.

\section{Ward Identities in Supersymmetric Quantum Mechanics}

We have tested these ideas by simulating the model given by the
action $S_{(1)}$
using a potential of the form given in 
eqn.~\ref{pot} which contains a mass term $m$ plus single 
coupling $g$. Details of our accelerated HMC algorithm are
given in \cite{alg}. The results were obtained for lattice
parameters $m=0.25$, $g=0.0625$
on an $L=4$ site lattice and utilized $10^8$ HMC trajectories. 
Since the dimensionless interaction
strength is given by the parameter $g/m^2=1$ this choice of
parameters corresponds to a strongly coupled theory on a
coarse lattice. As such it should easily reveal any symmetry
breaking in the lattice theory.
We examined the first non-trivial Ward identities corresponding to
the expectation values $<\delta_{(1)} O_1>$ and
$<\delta_{(2)} O_2>$ for $O_1=x_i\psib_j$ and $O_2=x_i\psi_j$.
These take the form
\begin{eqnarray*}
\left<x_iN^{(1)}_j\right>+\left<\psi_i\overline{\psi}_j\right>&=&0\\
\left<x_iN^{(2)}_j\right>+\left<\overline{\psi}_i\psi_j\right>&=&0\\
\end{eqnarray*}
The results are shown in tables 1 and 2. For an exact Ward identity
the sum of the bosonic and fermionic contributions across any row
should cancel. 
\DOUBLETABLE{
\begin{tabular}{||l|l|l||}
\hline
t  & $<x(0)N^{(1)}(t)>$      &$<\psi(0)\overline{\psi}(t)>$      \\\hline
0 & $0.8895(11)$ & $-0.8898(3)$ \\\hline
1 & $0.6152(10)$ & $-0.6155(3)$ \\\hline
2 & $0.4294(11)$ & $-0.4295(3)$ \\\hline
3 & $0.3024(11)$ & $-0.3028(3)$ \\\hline
\end{tabular}}
{
\begin{tabular}{||l|l|l||}
\hline
t  & $<x(0)N^{(2)}(t)>$      &$<\overline{\psi}(0)\psi(t)>$      \\\hline
0 & $-0.8895(11)$ & $0.8898(3)$ \\\hline
1 & $-0.3016(11)$ & $0.3028(3)$ \\\hline
2 & $-0.4294(11)$ & $0.4295(3)$ \\\hline
3 & $-0.6160(10)$ & $0.6155(3)$ \\\hline
\end{tabular}}
{Ward identity for $\delta_{(1)}$-symmetry at $g=0.0625$, $m=0.25$
and $L=4$}
{Ward identity for $\delta_{(2)}$-symmetry at $g=0.0625$, $m=0.25$
and $L=4$}
Since the action $S^{(1)}$ is invariant under the $\delta_{(1)}$-symmetry
it should be no surprise that the corresponding Ward identity is satisfied.
What is, on the surface, much more surprising is that the Ward
identity corresponding to the $\delta_{(2)}$-symmetry is also
satisfied to within the (small) statistical errors. This despite the
fact that the lattice action is not invariant under this symmetry.  
Of course this is just the result one would expect
on the basis of the arguments of the last section since $\delta_{(2)}$
generates an independent topological symmetry of the continuum
theory.
\TABULAR
{||l|l||}
{\hline
$\gamma$  &  $<S>$ \\\hline
0 & $3.9997(41)$ \\\hline
1 & $4.0003(39)$ \\\hline
2 & $4.0003(40)$ \\\hline
}
{$<S>$ vs $\gamma$ for $g=6.25$, $m=2.5$ and $L=4$}
Since the lattice actions 
$S_{(1)}$ and $S_{(2)}$ differ only by the cross term $C(g)$ and both
correspond to theories which
retain all the continuum topological symmetry for any coupling $g$ we
are led to conclude that perturbations about either lattice
theory via such a term will not affect topological observables.
The simplest
such observable is the partition function itself. To check this
we simulated the same model as before {\it except} we added the
cross term $C(g)$ to the action with arbitrary coupling $\gamma$.
\begin{displaymath}
S=S_{(1)}-\gamma C(g)
\end{displaymath}
Table 3. shows the
mean value of the action $<S>$ on an $L=4$ site lattice
using now $g=6.25$, $m=2.5$ for three values of the
coupling $\gamma$. Notice the data we show
here corresponds to even coarser lattices than before with the same
value for the interaction parameter $g/m^2$.
Here, the measured action includes 
contributions from the scalars and pseudofermions and, as detailed
in \cite{wz2}, can be shown to take the value
\begin{displaymath}
\left<S\right>=L+g\frac{\partial}{\partial g}Z\left(g\right)
\end{displaymath}
Thus, a partition function constructed from a BRST invariant
theory which
is independent of the coupling $g$ should yield $<S>=L$. At $\gamma=0$
and $\gamma=2$ (corresponding to lattice action $S^{(2)}$) this
is indeed the case. What is more surprising is that it appears to
be also true at $\gamma=1$ where the {\it classical} lattice action
is not
invariant under either the $\delta_{(1)}$ or $\delta_{(2)}$ symmetries.
We have checked this result for other values of $\gamma$ and
for different masses and couplings with identical
results.
To reinforce this point we show in tables 4 and 5 the same
two Ward identities as before computed in the $\gamma=1$ ensemble.
\DOUBLETABLE{
\begin{tabular}{||l|l|l||}
\hline
t  & $<x(0)N^{(1)}(t)>$      &$<\psi(0)\overline{\psi}(t)>$      \\\hline
0 & $0.23102(37)$ & $-0.23110(6)$ \\\hline
1 & $0.05367(15)$ & $-0.05334(3)$ \\\hline
2 & $0.01251(14)$ & $-0.01232(1)$ \\\hline
3 & $0.00263(11)$ & $-0.002848(3)$ \\\hline
\end{tabular}}
{
\begin{tabular}{||l|l|l||}
\hline
t  & $<x(0)N^{(2)}(t)>$      &$<\overline{\psi}(0)\psi(t)>$      \\\hline
0 & $-0.23102(38)$ & $0.23110(6)$ \\\hline
1 & $-0.00265(11)$ & $0.002848(3)$ \\\hline
2 & $-0.01251(14)$ & $0.01232(1)$ \\\hline
3 & $-0.05365(15)$ & $0.05334(3)$ \\\hline
\end{tabular}}
{Ward identity for $\delta_{(1)}$-symmetry at $g=6.25$, $m=2.5$
$L=4$ and $\gamma=1.0$}
{Ward identity for $\delta_{(2)}$-symmetry at $g=6.25$, $m=2.5$
$L=4$ and $\gamma=1.0$}
Again both Ward identities are satisfied within statistical errors.
Thus our numerical results
lend strong support to the results of section~\ref{formal}

\section{Complex Wess-Zumino Model in Two Dimensions}

These arguments can be extended to the complex Wess Zumino model with $N=2$
supersymmetry in two dimensions. A lattice formulation of this
model based on a discretization of the Nicolai map was studied
in \cite{niclat} and \cite{ES83}. More recently the
Nicolai map was extended to Ginsparg-Wilson fermions in \cite{gins}.
In \cite{ham}
a Hamiltonian approach was used to investigate the
same model. 

In \cite{wz2} we showed that this
model can be derived from the simple model eqn.~\ref{eq1} if one allows
the original index to represent both a lattice coordinate in two dimensional
Euclidean space and a two-component internal degree of freedom. 
The resulting theory contains Dirac fermions coupled to a complex
scalar field $\phi$ and in the continuum possesses $N=2$ supersymmetry.
This formulation of the model has the merit of exhibiting
clearly its topological character (the topological nature of this
model in the continuum is discussed in \cite{rev} and references therein)

In \cite{wz2}
we showed that there were four possible choices for the
scalar lattice action 
\begin{displaymath}
S\left(\phi\right)=\frac{1}{2}\eta^{(a)}\overline{\eta^{(a)}}
\end{displaymath}
each of which exhibited an exact fermionic symmetry and corresponded to
four inequivalent local Nicolai maps $\eta^{(a)}\left(\phi\right)$
with $a=1\ldots 4$. Each of these fermionic
symmetries is now seen to be a BRST symmetry corresponding to
four different quantizations of an underlying complex scalar
field theory possessing a local shift symmetry. The distinct Nicolai
maps simply correspond to different gauge choices for the scalars
possessing the {\it same} Fadeev-Popov determinant. The explicit forms of
these maps are (see \cite{wz2})
\begin{eqnarray*}
\eta^{(1)}&=&D^S_z\overline{\phi}+W^\prime(\phi)\nonumber\\
\eta^{(2)}&=&D^S_z\overline{\phi}-W^\prime(\phi)\nonumber\\
\eta^{(3)}&=&D^S_z\phi-W^\prime(\overline{\phi})\nonumber\\
\eta^{(4)}&=&D^S_z\phi+W^\prime(\overline{\phi})
\end{eqnarray*} 
They come in two groups of two corresponding to complex conjugation of
the scalar field $\phi$. Notice that here we are again using the
notation of section~\ref{susyqm} in which kinetic terms are
written in terms of symmetric difference operators and
wilson masses are added into the potential terms. Since $r=1$
this is entirely equivalent to the block language of
forward and backward derivatives and local potential
terms. The actions corresponding to the first two
of these maps differ only by a cross term of the form
\begin{displaymath}
C_1=2Re\left(D^S_z\overline{\phi}W^\prime(\overline{\phi})\right)
\end{displaymath}
Similarly the actions corresponding to the third and fourth Nicolai maps
would differ only by another cross term of the same form but with $\phi\to
\overline{\phi}$
\begin{displaymath}
C_2=2Re\left(D^S_z\phi W^\prime(\phi)\right)
\end{displaymath}

As before we can attempt to derive these lattice models
by blocking the continuum field theory. To do this we need 
generalizations to two dimensions of the lattice metric and blocking functions. 
The following choices appear to suffice (for simplicity we parametrize
the bosons in terms of two real fields here)
\begin{displaymath}
g(\sigma_1,\sigma_2)=diag\left(e_\beta^2(\sigma_1),e^2_\beta(\sigma_2)\right)
\label{2dmetric}
\end{displaymath}
The components of this diagonal matrix are just the squares of the
functions $e_\beta(t)$ introduced in section~\ref{formal}, each
now being a function of the corresponding coordinate $\sigma_i$.
Similarly the two-dimensional blocking function 
$B^{\rm 2d}_\beta(x)$ can be taken to be just the product of 
one-dimensional blocking functions, for example,
\begin{displaymath}
B^{\rm 2d}_\beta(\sigma_1,\sigma_2)=B_\beta^-(\sigma_1)B_\beta^-(\sigma_2)
\end{displaymath}

By following the same procedure as for supersymmetric quantum mechanics
we can derive a lattice action by blocking out of the continuum
a topological theory built from any 
of the Nicolai maps detailed above. Furthermore, topological observables
such as the Ward identities associated with
any of these continuum BRST symmetries should not depend on the couplings to
any cross terms of the form of $C_1$ or $C_2$
since such operators interpolate between equivalent topological field
theories.
Thus we predict that
the following simple lattice action will yield exact Ward identities
for all four symmetries at the quantum level even though it possesses
{\it none} of these symmetries at the classical level
\begin{equation}
S=\sum_{z}\frac{1}{2}\left(D^S_z\overline{\phi}D^S_{\overline{z}}\phi
+W^\prime(\phi)W^\prime(\overline{\phi})\right)+\ln \det{M}
\label{wz_nocross}
\end{equation}
where $M$ is the fermion operator corresponding to, for example, the
choice $\eta^{(1)}$ (the determinant of $M$ is independent of which map is
used). Notice 
\EPSFIGURE{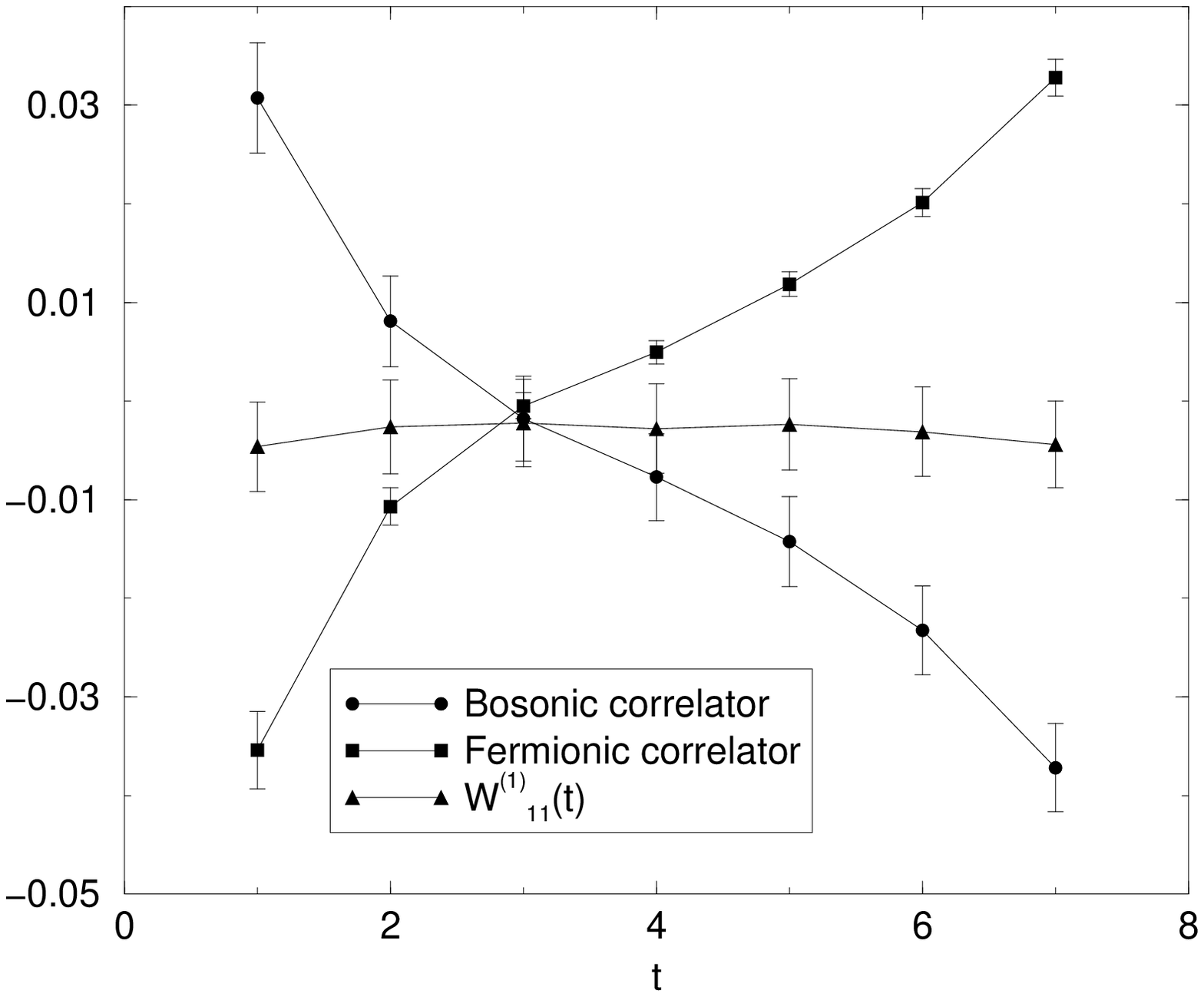,width=11cm}{Ward identity corresponding to $\delta_{(1)}$-symmetry
with $m=0.625$, $g=0.625$ and $L=8$}
\EPSFIGURE{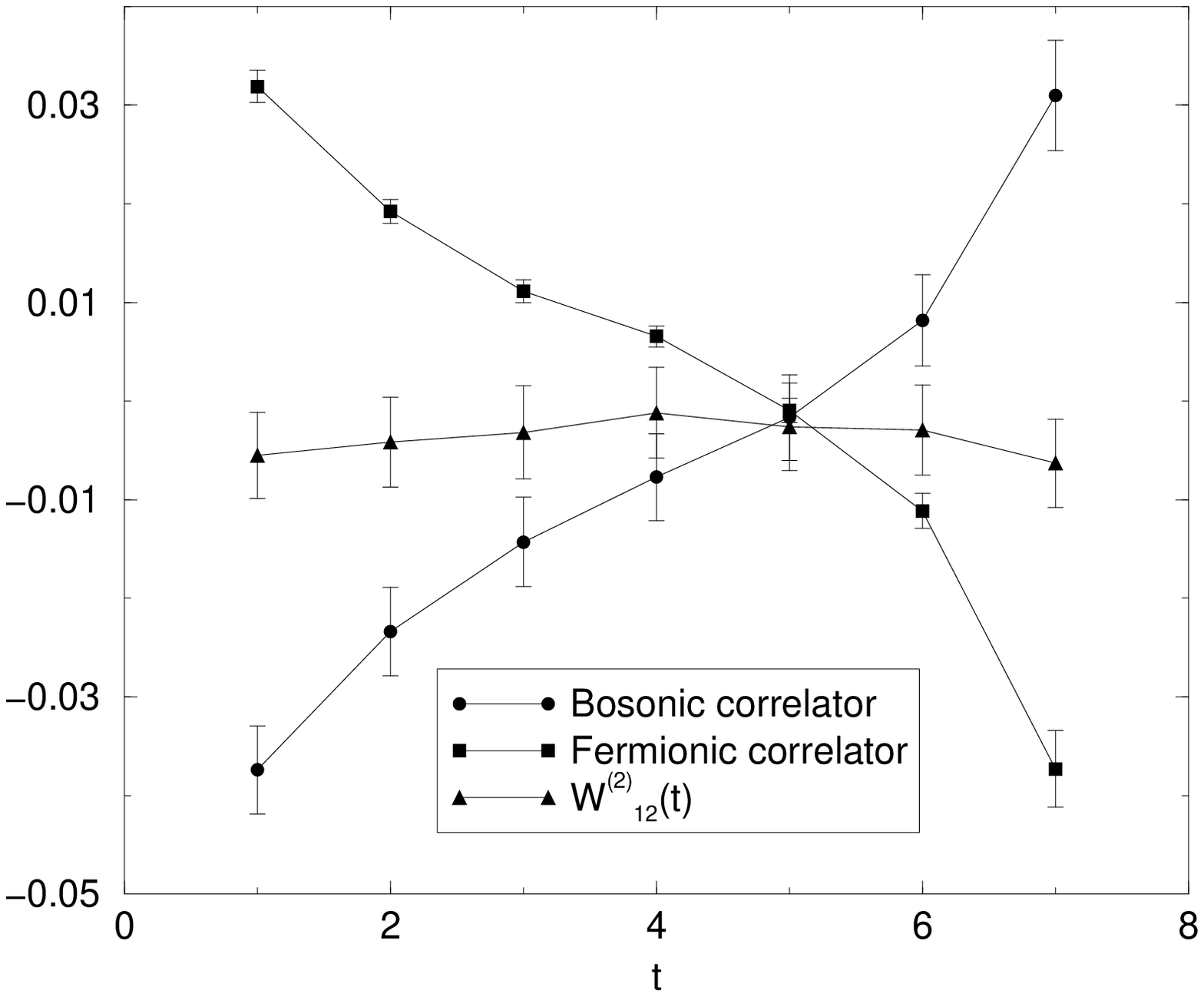,width=11cm}{Ward identity corresponding to $\delta_{(2)}$-symmetry
with $m=0.625$, $g=0.625$ and $L=8$}

We have checked these conclusions by measuring the four simplest
non-trivial Ward identities corresponding to these BRST symmetries
for the action given by eqn.~\ref{wz_nocross} on an $8\times 8$ site
lattice with $g/m=1$ and lattice mass $m=0.625$. 
The detailed form of these was derived in \cite{wz2}. For completeness
we list them again here

\begin{eqnarray*}
0&=&\left\langle\psi_i^\alpha \psib_j^\beta\right>+\left<N_j^\beta
x_i^\alpha\right\rangle\nonumber\\
0&=&\left<i\gamma_3^{\alpha\gamma}\psi_i^\gamma\psib_j^\beta\right>+
\left<i\gamma_3^{\beta\gamma}\overline{N}_j^\gamma x_i^\alpha\right>\nonumber\\
0&=&\left\langle\gamma_1^{\alpha\gamma}\psib_i^\gamma\psi_j^\beta\right\rangle+
\left\langle Q_j^\beta x_i^\alpha\right\rangle\nonumber\\
0&=&\left\langle\gamma_2^{\alpha\gamma}\psib_i^\gamma\psi_j^\beta\right\rangle+
\left\langle i\gamma_3^{\beta\gamma}\overline{Q}_j^\gamma
x_i^\alpha\right\rangle
\end{eqnarray*}

Notice that these expressions involve not the original complex boson fields
$\phi,\eta^{(i)}$ but their real counterparts eg. $\phi=x_1+ix_2$ and
$\eta^{(1)}=N_1+iN_2$ with similar expressions for $\eta^{(2)}$, $\eta^{(3)}$
and $\eta^{(4)}$ in terms of $\overline{N}_\alpha$, $Q_\alpha$ and
$\overline{Q}_\alpha$ respectively where $\alpha=1,2$. Fig.~1 shows a plot
of the $(11)$-component of the first Ward identity corresponding to the
$\delta_{(1)}$-symmetry. For simplicity all the correlators
we exhibit correspond to timeslice averaged fields. 
Both bosonic and fermionic contributions are
shown together with the combination yielding the Ward identity. It is clear
that the latter is satisfied within statistical errors even though the
starting lattice action given by eqn.~\ref{wz_nocross} does not possess
this symmetry exactly. We find this situation to be true for
all the Ward identities -- as a further example, fig.~2 shows the $(12)$-component of the
2nd Ward identity. Again, we have checked these conclusions for
a variety of lattice parameters and sizes with the same result.

In \cite{finite} it was argued that the presence of a local Nicolai map
in the lattice model played no essential role in determining the
renormalization properties of the theory. However, the
existence of a local Nicolai map can now be viewed as
a consequence of the topological character of the
continuum theory. Thus we would claim that the benign
U.V behavior of the lattice model in eqn.~\ref{wz_nocross}
is intimately connected to
the existence of an exact Nicolai map in the associated continuum
model.

\section{Discussion}

It is well known that certain low dimensional non-gauge models can be
discretized in such a way as to preserve a single parameter fermionic
symmetry. In two cases - supersymmetric quantum mechanics
and the complex Wess-Zumino model in two dimensions we show
that this lattice invariance (and the
associated existence of a local Nicolai map) follow from an underlying
topological structure in the continuum field theories. Indeed
certain Ward identities of the supersymmetric model just correspond
to trivial topological observables in the topological field theory.
Furthermore, we argue that these lattice models can be arrived at by
blocking out of the continuum in a carefully chosen background
geometry. Such a procedure will
generate a {\it perfect} lattice action for the computation of topological
observables and hence guarantees an absence of
cut-off effects in the lattice Ward identities. 
For the models considered here which contain two and four
topological symmetries respectively this connection of the lattice model
to the continuum guarantees that {\it all} the supersymmetric Ward
identities are satisfied exactly on the lattice. This may even be true
for lattice actions which are {\it not} classically invariant under the
some of the supersymmetries. 

The crucial ingredient in
topological field theories of this type
is that they contain a scalar nilpotent charge arising
from BRST quantizing an underlying bosonic theory. In terms of the
original supersymmetry this charge is obtained by taking particular
linear combinations of the supercharges. 
This latter procedure requires that the original theory
possess extended supersymmetry. 
The question that remains is how much of this structure survives
in higher dimensions and in the presence of gauge symmetry.
The procedure to obtain a nilpotent charge from a set of supercharges
is termed {\it twisting} and is a well known method to obtain
topological field theories from theories with extended supersymmetry
in higher dimensions \cite{witten}. Indeed, Donaldson-Witten
theory was the first topological field theory to be constructed
and corresponds to twisting $N=2$ super
Yang Mills theory. In general, in higher dimensions, only a fraction
of the original supersymmetries can be reinterpreted as yielding
BRST charges and it is only this fraction that we can hope to
preserve on the lattice. Nevertheless, it is tempting to try and
use this topological field theory interpretation to
construct lattice models containing a residual element of
supersymmetry. The procedure of blocking out of the
continuum in the presence of a carefully chosen ``lattice'' metric may
prove very fruitful in this regard. 

While we think that this approach deserves 
further study it will clearly be problematic - the lattice theory,
to duplicate the structure of the continuum theories, must contain
non-compact fields, and so such theories will necessarily
break gauge invariance at the classical level. To utilize
this formalism it will then be necessary to show that the existence of
this residual supersymmetry can constrain the radiative corrections
sufficiently to eliminate dangerous gauge-violating
counterterms. Notice additionally, that the presence of
more than one topological symmetry may be necessary in order to establish an
explicit connection between the lattice and continuum theories.

Additionally, local Nicolai maps are known to
exist for a number of
other models \cite{nicsym} not all of which have
$N=2$ supersymmetry. This is intriguing as it points to
a possible hidden topological structure in those models which, if
elucidated, may itself help with the problem of studying such 
models on lattices.

\acknowledgments
The author would like to thank the Niels Bohr Institute for hospitality
during the initial stages of this work and Poul Damgaard, in particular, for
numerous discussions on topological field theory. The numerical
data on the Wess Zumino model was provided by Sergey Karamov. The
author would also like to thank Joel Rozowsky for a careful
reading of the manuscript. This work
was supported in part by DOE grant DE-FG02-85ER40237.

\end{document}